\input{aipcheck}

\documentclass[
    ,final            
  ]
  {aipproc}

\layoutstyle{6x9}

\begin{document}

\title{Radio to TeV radiation initiated by termination of 
hadronic jets from microquasars in the ISM}

\author{V. Bosch-Ramon}{
  address={Departament d'Astronomia i Meteorologia, Universitat de Barcelona, 
  Av. Diagonal 647, E-08028 Barcelona, Catalonia, Spain}
}

\author{F. A. Aharonian}{
  address={Max-Planck-Institut fur Kernphysik,
Saupfercheckweg 1, Heidelberg, 69117, Germany}
}
\author{J.M. Paredes}{
  address={Departament d'Astronomia i Meteorologia, Universitat de Barcelona, 
  Av. Diagonal 647, E-08028 Barcelona, Catalonia, Spain}
}

\begin{abstract} 
Microquasars (MQs) are potential candidates to produce a
non-negligible fraction of the observed galactic cosmic rays. The protons
accelerated at the jet termination shock interact with the interstellar medium and
may produce extended emission detectable at different energy bands through 
several processes:
neutral pion-decay produce high-energy and very high-energy gamma-rays, 
secondary electrons produced by charged pion-decay generate 
synchrotron and bremsstrahlung emission. In addition, the jets of MQs themselves
are likely  sources of gamma-rays. We discuss about the association between the
intrinsic and the indirect emission coming from these objects. 
\end{abstract}

\maketitle


\section{Introduction}  

MQs are X-ray binary systems with relativistic jets or outflows emitting non-thermal
radio emission (\cite{Mirabel99}). Although the mechanisms of jet generation are
still under discussion (\cite{Meier03}), it is known that these outflows are
magnetized, and they contain and likely accelerate relativistic particles that 
produce the observed extended radio structures. Moreover, the radio band seems to be not
the only energy range which these jets emit at. The likely association between MQs
and gamma-ray sources based on observational (for LS~5039, see \cite{Paredes00} 
and \cite{Aharonian05}; for
LS~I~+61~303, see \cite{Massi04}) and theoretical grounds (\cite{Atoyan99}, \cite{Romero03},
\cite{Bosch05a}) point strongly to these objects as very high energy cosmic ray
accelerators, not only of leptons, but perhaps also of hadrons
(\cite{Migliari02}, \cite{Bosch05b}). 

If relativistic hadrons leave the region of acceleration, they can interact further
away with high density regions of the ISM (e.g. clouds), and the electromagnetic
spectrum generated in such interactions would present features depending on the
propagation effects (\cite{Aharonian96}) linked to the diffusive medium as well as
on the properties of the cloud and the jet itself. We have explored here which
consequences in the observed spectrum, as well as concerning the gamma-ray imaging, would
arise from a region where a gamma-ray emitting microquasar jet is interacting with a
nearby cloud.

\section{Emission from regions containing MQ and clouds}

We assume that a significant amount of protons, accelerated in the
region where the jet ends, are released interacting further away 
with the ISM (see \cite{Heinz02}). These
protons diffuse through the ISM with a diffusion coefficient homogenous in spaces 
that follows a power-law in energy. Due to propagation effects, the outcomes of the
interactions between the protons released from the jet and cloud hydrogen nuclei can
differ strongly depending on the age, the nature (impulsive or continuous) of the
accelerator and the distance between this and the cloud. The proton-proton
collisions within the cloud lead to the creation of neutral and charged pions. Then,
neutral pions will decay to gamma-ray photons while charged pions will decay to e$^-$ and
e$^+$. These secondary particles can produce significant levels of synchrotron
(from radio frequencies to X-rays) and Bremsstrahlung emission (from soft gamma-rays
to TeV range), and generally with much less efficiency, inverse Compton emission
through interaction with the ambient infrared photons. In Table~\ref{tab:a}, we
present parameter values adopted concerning relevant physical quantities for a
middle size molecular cloud, the ISM and a jet of a MQ. Also, in Figs.~1 and 2, we
show the computed broad-band spectral energy distribution (SED) of the radiation
coming from the bombarded cloud for the continuous and the impulsive case respectively. 

\begin{table}
\begin{tabular}{cc}
\hline
Parameter & Value \\
\hline
diffusion coefficient normalization constant at 10 GeV & $10^{27}$~cm$^2$~s$^{-1}$ \\
diffusion power-law index & 0.5 \\
ISM medium density & 0.1~cm$^{-3}$ \\
cloud density & 10$^4$~cm$^{-3}$ \\
mass of the high density region/cloud & $3\times10^4~M_{\odot}$ \\
magnetic field within the cloud & $5\times10^{-4}$~G \\
IR radiation energy density within the cloud & 10~eV~cm$^{-3}$ \\
power-law index of the high energy protons & 2 \\
maximum energy of the high energy protons & $10^5$~GeV \\
kinetic luminosity of accelerated protons in the MQ jet & 10$^{37}$~erg~s$^{-1}$ \\
kinetic energy of accelerated protons in the impulsive ejection & 10$^{48}$~erg \\
distance between the MQ and the cloud & 10~pc \\
\hline
\end{tabular}
\caption{Adopted parameter values}
\label{tab:a}
\end{table}

\begin{figure}
\includegraphics[angle=0, width=10cm]{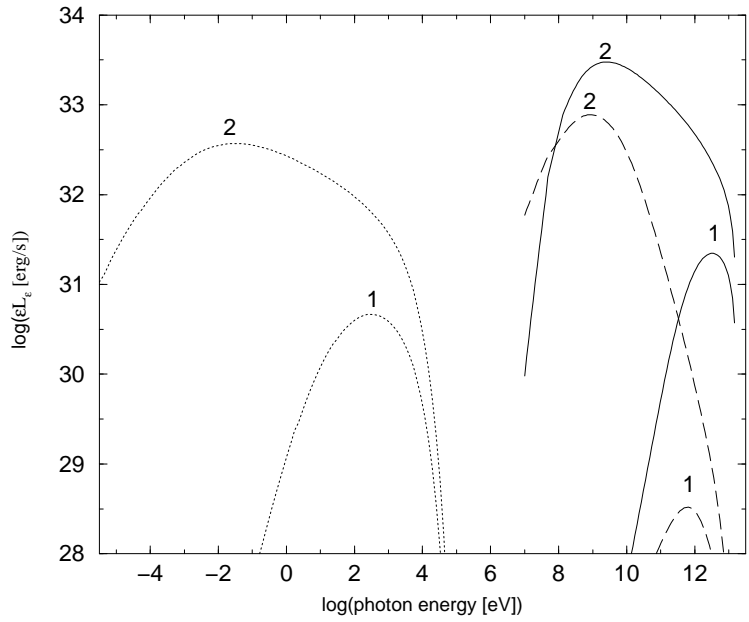}
  \caption{SED for a continuous MQ from radio to very high-energy gamma-rays at two
different ages: t=100 yr (1), t=10000 yr (2). It is plotted the neutral pion-decay 
gamma-rays (solid line), Bremsstrahlung (dashed line) and synchrotron emission
(dotted line).}
\end{figure}

\begin{figure}
\includegraphics[angle=0, width=10cm]{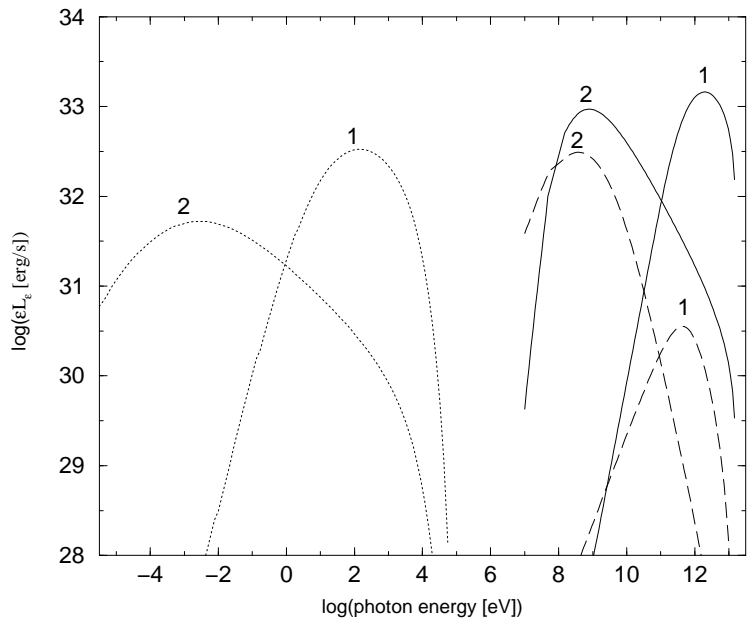}
  \caption{The same as in Fig. 1, but for an impulsive microquasar.}
\end{figure}

\section{Discussion}

From the figures shown in the previous section, it is seen that a cloud interacting with the
cosmic rays released from jet of a MQ could be detected  by ground-based cherenkov
telescopes, and the same applies at high-energy gamma-rays for satellite-borne instruments.
However, as noted above, MQs are also intrinsic sources of very high-energy gamma-rays,
detected already at TeV energies, and likely sources of high-energy gamma-rays. Theoretical
studies show also that jets could be emitting emission in the whole spectral range
(\cite{Paredes05}). This implies that these objects, when observed with good sensitivity and
angular resolution instruments, could appear as compact sources with nearby 
extended hot spots, being the spectra of both related to each other and to the separation
distance and age and nature of the accelerator. The discovery of such double sources would
provide of information about the physical properties of the clouds and the MQs themselves.
For instance, it would be possible to estimate the magnetic field and hydrogen density in the
cloud, the magnetic field, the accelerated particle spectrum and the relativistic hadronic
content in the jet, and finally the diffusion coefficient of the ISM. 

\section{Summary}

We have shown that jets of MQs can indirectly generate broad-band emission up to very high
energies through interactions between their hadronic content, accelerated in and released
from the jet, and the hydrogen nuclei of nearby molecular clouds. Due to the intrinsic nature
of these objects as gamma-ray emitters, the new ground-based Cherenkov telescopes and
satellite-borne gamma-ray instrumens could detect double sources in their fields of view: the
MQ itself and the jet proton bombarded high density ISM.



\begin{theacknowledgments}
V.B-R. and J.M.P. acknowledge partial support by DGI of the spanish Ministerio 
de Educación y Ciencia (MEC) under grant AYA2004-07171-C02-01, as well as additional 
support from the European Regional Development Fund  (ERDF/FEDER). V.B-R. is 
supported by the DGI of the spanish MEC under the fellowship BES-2002-2699. 
\end{theacknowledgments}



\bibliographystyle{aipproc}   

\bibliography{sample}

\IfFileExists{\jobname.bbl}{}
 {\typeout{}
  \typeout{******************************************}
  \typeout{** Please run "bibtex \jobname" to optain}
  \typeout{** the bibliography and then re-run LaTeX}
  \typeout{** twice to fix the references!}
  \typeout{******************************************}
  \typeout{}
 }



\end{document}